**Relativistic virtual worlds: an emerging framework**
Bradly Alicea
Michigan State University
East Lansing, MI
48824
bradly.alicea@ieee.org

**Abstract**

In this paper, I will attempt to establish a framework for representation in virtual worlds that may allow for input data from many different scales and virtual physics to be merged. For example, a typical virtual environment must effectively handle user input, sensor data, and virtual world physics all in real- time. Merging all of these data into a single interactive system requires that we adapt approaches from topological methods such as *n*-dimensional relativistic representation. A number of hypothetical examples will be provided throughout the paper to clarify technical challenges that need to be overcome to realize this vision. The long-term goal of this work is that truly invariant representations will ultimately result from establishing formal, inclusive relationships between these different domains. Using this framework, incomplete information in one or more domains can be compensated for by parallelism and mappings within the virtual world representation. To introduce this approach, I will review recent developments in embodiment, virtual world technology, and neuroscience relevant to the control of virtual worlds. The next step will be to borrow ideas from fields such as brain science, applied mathematics, and cosmology to give proper perspective to this approach. A simple demonstration will then be given using an intuitive example of physical relativism. Finally, future directions for the application of this method will be considered.

**Introduction**

Virtual Environments have been a staple of science fiction and engineering innovation for decades, but are becoming an enabling technology in fields ranging from medicine to entertainment. Yet there is still much work to be done in the area of virtual representation. This is particularly true as applications become increasingly unconventional. Insights from a wide range of scientific fields might be able to help out conceptually by tying together human behavior and virtual worlds using concepts from relativistic approaches.

Why relativism? Insights from virtual world control

To understand why high-dimensional, relativistic spaces could be useful in application to virtual environments, we must consider where the research on control of virtual environments is headed. In the early years of virtual environments, control of virtual worlds was limited to simple interaction devices such as keyboards, computer mice, and joysticks. These devices provided sparse, discrete information to a virtual environment [1]. In recent years, input variables such as kinematics, pressure, and video segmentation have been incorporated into an increasingly realistic scheme of image rendering and simulated physics (Figure 1). At the same time, the individual who controls a virtual environment ultimately uses their brain and body to realize action in this synthetic world. Early in the development of virtual environments, it was not considered necessary to couple action in the virtual world and physical control [2]. As both the range of applications and immersion capabilities of virtual worlds continue to increase, this is quickly becoming a necessity. Two elements of human action will become critical to virtual world control: embodiment and neural control. Together, they form the basis for representation.

Representation as a "basis" function

The consequence of taking inputs from an individual possessing their own complex set of representations and merging it with a series of digital spaces are far-reaching [3]. To understand how high-dimensional, relativistic virtual spaces would be beneficial, it is important to distinguish between a purely technological representation and a cognitive embodied representation. While the former provides us with a framework for simulation, animation, interaction, the latter type of representation treats space, objects, and relationships are treated as a mental phenomenon.

Representation and haptic phenomena. This disparity can be understood using the example of size-weight illusion and dynamic touch [4]. The size-weight illusion involves presenting a specific virtual object such as a barbell. When the user picks up the barbell, the accompanying weight and pressure cues are missing. The underlying perceptual dynamics of this effect has been characterized by Weber's Law [5], but the relationship between the architecture of a virtual barbell and a neural representation of a seeing, touching, and lifting a barbell is not well understood.

Decoupling representations. Another example of where these two types of representation interact is in the area of selectively decoupling perception and action. Decouplability [6, 7] is a term used in the embodiment literature for characterizing the ability of using inner states to guide behavior in the absence of represented features. This plays a role in cognitive functions such as visual attention, where the environment is sampled discretely, but continuous action is extrapolated by what the brain expects to happen [8]. This can be exploited by a relativistic virtual world in ways that current approaches to virtual design do not provide. As we have just seen, virtual worlds have many unique properties that make a straightforward mapping of physical input to digital commands rather difficult. Key among these properties is the inhomogeneous nature of virtual representation, particularly with respect to different types of input.

**Elements of a Relativistic Topology**
There are three elements that define a relativistic topology, which will be discussed on a largely conceptual basis herein. The first of these is isomorphism between digital and representational spaces which will be discussed further in Section 2.1. The second involves relativistic translation between multiple virtual and real spaces. By translating between real and digital spaces without the requirement of isomorphism may allow us to go beyond realism in virtual worlds within a systematic framework. This desire is based partially on the need for "fantasy" world physics in certain applications, and partially on cases where the level of human agency has no relationship with increasing amounts of realism [9]. Thirdly, we must be able to simulate physical attributes of these relativistic spaces, or map between them in real time. This includes both well- and lesser-characterized attributes such as light, sound, and non-Newtonian physics.

Isomorphism
One way to characterize the relationship between real and virtual worlds is as classes of isomorphism [a]. Virtual and neural representations that are most tightly coupled are considered to be isomorphic. That is, each element of the virtual representation can be mapped to a corresponding element of the neural representation of that object in the real world. This can be demonstrated by the power of realism in virtual worlds: objects that have the proper physical and geometric characteristics are generally thought to be "usable" [10].

Properties of digital spaces. Gabor [11] has provided a conceptual framework called digital spaces for understanding transformations between different computational frames of reference in a purely digital system. In this definition, there are two criteria used to define an isomorphic mapping: translatability and reversibility.

Translatability refers to the ability to translate between coordinate systems, and has immediate application to action that is distributed between the real and virtual worlds.

Translatability is also applicable to systems distributed over multiple contexts, such as mobile devices or different geographical locations. Reversibility is the ability to retrieve and restore information from a data structure. Traditionally, this has been accomplished through data compression schemes [12, 13]. However, this is becoming an emerging topic in the programmability of matter [14]. In this and similar contexts, relativistic data structures might allow for modeling physical objects at multiple scales of complexity.

Taken together, translatability and reversibility may also allow us to create alternate physical worlds *in silico*. Mappings between geometric spaces representing the real and multiple virtual worlds might be implemented in such a way so that improving the immersive qualities of this virtual world without simply making it a matter of increasing realism [15].

Relativistic translation
One example of this can be shown by translating pointing and reaching in two dimensions to a higher-dimensional context. Pointing and reaching has a well-defined set of neural and behavioral characteristics [16]. However, the mapping between real and virtual worlds shows the difficulty posed in virtual world control. Moving a cursor in a 2-D virtual environment with an arm controlling a mouse in an unpredictable 3-D environment is not the way the nervous system moves the arm in a real world environment with physical objects [17, 18]. Thus, the characterization of human movement herein has been made unduly linear by mapping complex kinematic inputs to a sparse representation [19].

Acceleration and deceleration. Two physical parameters that are critical to this mapping are acceleration and deceleration, which are actually second-order derivatives of position. In other words, acceleration and deceleration are not simply parameters based on first-order differentials. Besides changes in acceleration, the arm driving a mouse can produce effects that are third, fourth, fifth, and sixth-order derivatives of position during movement [20]. Relativistic virtual worlds, with their emphasis on physical constants, may be able to deal with this issue more efficiently than traditional approaches.

Higher-order dynamics. Given the dimensionality reduction of traditional virtual world design, relativistic spaces might allow us to capture the higher-order dynamics that modulate/influence arm movements towards a target. Upon transfer to the real world, this becomes quite important. Examples of this include the study of movement disorders for the design of assistive devices and the control of variable loads such as pouring water into a cup.

Two components of relativistic virtual worlds
To achieve relativism, two types of virtual space are required: absolute but overlapping spaces, and true relativistic spaces. The way in which these types of spaces work together will be application-dependent.

Absolute overlapping vs. true relativistic space. One type of relativistic space, as in all virtual worlds, involves absolute representation. However, in the relativistic scheme, these absolute representations can overlap. This may provide a means for relativistic representation in the spatial domain. True relativistic spaces, on the other hand, serve as a mechanism for dynamic control. Such spaces can be used as placeholders in order to introduce temporal lag. In other cases, multiple spaces are used to simulate action at many different rates, the mapping between which may allow for users to travel between dimensions.

Simulating physical attributes

While the realistic simulation of physical objects is an active area of research with many good solutions available [21], physical simulation in the context of interfacial qualities, surface reaction forces, and phases of matter [22] have yet to be addressed. Specifically, the interfacial qualities between phases of matter, compliance of the surface, and properties of friction, adhesion, and shear should all be important dimensions of a virtual representation currently underspecified in virtual worlds. In this section then, it will be shown how a relativistic approach might play a role in resolving both the underappreciated and unpredictable properties of these physical phenomena.

Bouncing rubber ball example.

I will now briefly discuss the implications of each type of space used for the purpose of simulating a bouncing rubber ball. The first features of any "relativistic" physical object will involve modeling its energy, mass, and light reflecting properties (see Section 6.2). In a conventional virtual world, these properties would be determined by a set of parameters and governed by their values given interaction with the object in question [23].

In a relativistic space, energy and mass are related using a common frame of reference (see Section 6.3). In terms of the representation, mass, energy, light, and an environmental constant are all held in different spaces. The secondary properties of the ball (material properties, feel, etc) are held in secondary maps which are contingent on behavior of spaces representing the physical laws. The operators between these spaces allow us to change the nature of the bouncing ball depending on who is viewing it and where in the virtual universe they are.

Interaction in a relativistic virtual universe. Interaction can also be mediated by relativistic virtual spaces. This can be understood by One prediction of relativity is that objects traveling at constant velocity will exhibit variable properties relative to the observer. According to [24], it is argued in special relativity that observers can describe physical phenomena in their own times and spaces. This capacity for selective geometric perspective could be useful in terms of providing individual users with unique experiences, while at the same time maintaining a universal set of environmental constants.

Another prediction of relativity involves the constant curvature of space [25]. Greene [26] provides an example of a fly that lives on a garden hose. The fly is unaware of the curvature inherent in the garden hose surface, much as we have no intuitive sense that our earth and universe has a curvature to its topology. A third feature of relativity is related to the constancy of light. One property of light is that there is a correspondence between the geometry between light and matter [27]. As discussed in the bouncing rubber ball example, physical attributes should be invariant with respect to multiple dimensions [28]. In relativistic virtual spaces, light may act as a way to map between different dimensions and unify these spaces into a single context.

Toward multidimensional objects. Ultimately, these examples lead us to multidimensional physical modeling. This may include conventional three-dimensional objects embedded in a relativistic context. Multidimensional modeling may also be useful in modeling complex, nonlinear phenomena such as non-Newtonian fluids. The inertial and other properties of force are highly nonlinear and even stochastic. The challenge will be to model such materials in a way so as to capture their true nature without losing the ability to interact with the virtual model in real-time.

More generally speaking, there are two advantages to multidimensional parallel modeling. For example, we can have "warped" versions of the spatial and temporal aspects of an object or scenario (see Figure 2). There will be nonlinear relations with regard to dimensionality. The second is the ability to create nonlinear force fields. This provides different dimensions of different types. However, the domains of space and time and the dimensions they embody can be relatable and addressable even after warping.

**Towards higher-dimensionality**

In this paper, the $n$-dimensional nature of relativistic data structures will be made relevant to two representation problems: alternate virtual universes and the rendering of surfaces, interfaces, and materials. However, one might ask why we need high-dimensionality representations in virtual environments. Higher-dimensional phenomena are common in the natural world. In computational domains such as visualization [29], higher-dimensional systems are often reduced to three dimensions. However, this does not mitigate the need to recognize contributions of higher-dimensional phenomena to simulations of the real world. There are two domains of prior research that could help us achieve relativistic representation in virtual worlds: cosmological and brain science. We will now turn to examples from each area of research and discuss how more advanced findings in these fields might contribute to the relativistic virtual space framework.

Alternate virtual universes

The proliferation of dimensions in contemporary physics is required to deal with the chaotic nature of the quantum world and distortions imposed by physical constraints. However, multiple dimensions can be a useful mechanism for virtual environments, as a system of dimensions might have myriad uses in simulating both realistic and fantastical scenarios [2, 26]. As with the distinction between biological and computational representation, there are further principles that can be extended from other domains of science. Cosmology and brain science each have provided natural examples of multiple dimensions and mapping between representations, respectively. Explicit connections between these areas and the framework in this paper are limited, but will serve to inspire future work.

Contributions of cosmology. Typically, phenomena that occur in a living room can be represented in four dimensions (three spatial dimensions and a temporal dimension). This is the most intuitive way to model the real world. However, it is not necessarily accurate. The Kaluza-Klein model of five-dimensional space opened up the possibility for multiple temporal dimensions currently embodied by string theory [25, 27]. In particular, string theory offers descriptions of 11-dimensional Calabi-Yau spaces that have been used to describe the nature of reality (see Figure 3, [27]). This serves simulated environments well, since multiple temporal dimensions can be used to both represent action or to serve as a placeholder.

Contributions of brain science. The brain provides us with a natural model for encoding, mapping, and addressing high-dimensional information. In particular, the hierarchical and laminar system in mammalian cortex provides inspiration for two mechanisms: multisensory integration and interregional communication [32, 33]. In Figure 4, we can see that in neural structures such as the thalamus and visual cortex, information can be encoded, indexed, and stored for retrieval. This information is mapped in ways that approximate both simple, linear and complex, nonlinear features of the environment.

Concrete examples in virtual worlds

In this section, I will introduce two examples of relationships between dimensions. These relationships can apply to both spatial and temporal dimensions. Orthogonal dimensions are those that exist in parallel but intersect linearly over a range of values. On the other hand, deformed relations are dimensions that intersect nonlinearly over a range of values.

Orthogonal dimensions. The first example involves orthogonal dimensions. Two dimensions (A and B) are orthogonal with respect to its physical and kinetic properties. The information in these dimensions is ideally independent in a statistical sense, and is linked solely through physical properties. The example in Figure 2 demonstrates how this independence provides separable sets of objects and scene properties. This separability is absolute with regard to geometric orientation, so that one object set does not interfere with the other.

Deformed relations between dimensions. The second operational example involves deformed relation between dimensions. In this case, dimensions A and B are convoluted with respect to one another. The convolution is representative of deformation between each dimension. Figure 2 shows how one dimension can be deformed, and so their relationship is also deformed. Maximal deformity results in dimensions that curl up like twisted strings: they can either curl up about their own axis or curl up with respect to each other.

Example #1: physical constants. As mentioned previously, there are several physical constants that serve as the basis for mappings in relativistic virtual worlds. As shown in Section 6.1, energy, mass, and light are all interrelated quantities. The coupling results in the intertwining of physical, kinetic, and component features of a given environment, without respect to its degree of realism.

Example #2: scaling vs. deformation. Isometry has been previously mentioned as a way to characterize realistic virtual representations. However, it is known from natural systems that there can be two effects when objects are increased in scale. Figure 5 demonstrates these two scaling regimes in natural systems. Isometry involves a self-similarity with regard to objects of increasing size, while allometry involves a nonlinear change with regard to changes in size (see Section 6.3).

In virtual worlds, isometry and allometry are proposed to be key features of a touch-based system and likewise in relativistic virtual worlds. As the interface or effector is scaled up with regard to size, the effects may or may not be linear. [37, 38]. This is an allometric effect, and is characteristic of touch perception that conforms to Weber's Law. However, it is worth considering that physical laws in relativistic virtual representations might also conform to this type of scaling law. This may be particularly true when scaling from a set of like objects to a heterogeneous environment.

**Future Application Domains**

There are many outstanding issues related to the design of virtual worlds with respect to building robust, complex, representations. However, the relativistic representations introduced here have several potential application domains.

Mixed-mode interfaces.
In the brain, multisensory integration provides a mechanism for enhanced information processing and perhaps even a basis for consciousness [39]. Using relativistic spaces to represent auditory, haptic, and auditory/haptic information [for example, see 32] might bridge the context gap between haptic and auditory modes of communication between the real and multiple virtual worlds. Other applications might involve the multisensory enhancement that occurs when tool use is integrated virtual world environment [40, 41]. In this case, spaces immediate to the user's body and where tool use occurs is qualitatively different from other virtual space, and thus has a relativistic component.

Phase transition behavior in materials.
In materials such as water, changes between phase (e.g. solid, liquid, gas) are characterized as first-order phase transitions with nonlinear, chaotic behavior [44]. Relativistic virtual worlds could represent this phenomenon by mapping each phase to a separate space, and then using a mapping scheme to simulate the phase transition as a stochastic process. A related phenomenon is the integration of time and topology in biological systems, particularly in the growth of surfaces [45].

Mass customization.
Multi-dimensional virtual worlds open the door for personalized simulations. However, as envisioned, the power of relativistic virtual worlds involves integrating simulated environments with real environments across contexts. This could be useful in augmented reality applications, where integrating virtual representations with moving real contexts is critical to coherent simulated experiences [11].

**Conclusions**
The main idea presented in this paper is that the topologically-based concept of relativism may be used to advance the state of virtual world representations. This technical advancement may be especially pronounced for systems that involve camera-based tracking, accelerometer control, or context-sensitive mobile devices. This key component of relativistic virtual worlds, their component spaces, and input device dynamic information is *motion*, particularly when both lower- and higher-derivative descriptions of position are required in the same application. Further development of this method will involve more advanced mathematical modeling along with greater integration of biological mechanism and human factors engineering. While this is an ambitious call to arms, the payoff is potentially enormous.

**Methods**

This section specifies the methods that might be used to create relativistic virtual worlds. While the technical details have yet to be worked out, I will provide a rough sketch of the expected technical challenges.

Parallel computation
Multiple dimensions in virtual worlds can be realized using a parallel computing approach [46, 47]. Ideally, each dimension could be simulated on a different core. Each dimension consists of a frame of reference and registration of objects with regard to either space or time, depending on the type of dimension.

Equations for relativistic models
In this section, a series of elementary equations will be introduced as an intuitive example of how physical parameters are interrelated in a relativistic virtual world. These equations are based on the famous equation summarizing special relativity and the equations of motion according to classical mechanics.

The equation which makes the connection between light, objects traveling through space, and their potential energy is stated as

$$E = MC^2 \quad [1]$$

This formulation, when used in the context of virtual worlds, implies an energy function (E), which can be defined by considering the components of mass (M) and the global constant (C).

As an essential part of this formulation, mass is related to force in the following manner

$$M = \frac{F}{A} \quad [2]$$

where F is force and A is acceleration. Acceleration (A) is the second derivative of position (P''). Thus, equation 2 can be restated as

$$M = \frac{F}{P''} \quad [3]$$

and equation 1 becomes

$$E = \left(\frac{F}{P''}\right) C^2 \quad [4]$$

In some cases, equation [4] can also be treated as a dynamical system. In general, this formulation allows us to treat individual object trajectories inside the virtual world and movement from the non-virtual world to be given the same currency.

Special role of light in relativistic virtual worlds

Light also serves a unifying function in relativistic virtual worlds. A typical light shading equation is shown in [24], and describes how light is treated in a conventional virtual world with rendered objects. Light is emitted from a point source (x), while illumination is a function of two key variables: angle of a light source relative to available surfaces, and color/reflectance properties of a given surface within a finite volume. In a relativistic virtual world, it is envisioned that light would be a more universal physical parameter that provides a force-feedback like response, enabling communication between virtual dimensions and between the virtual world and the user.

Allometric scaling

Allometric growth is defined by the following equation

$$y \sim f(x)^\alpha \quad [5]$$

where allometric relationships are typically an $\alpha$ value close to or beyond +/- 1.0.

## Figures

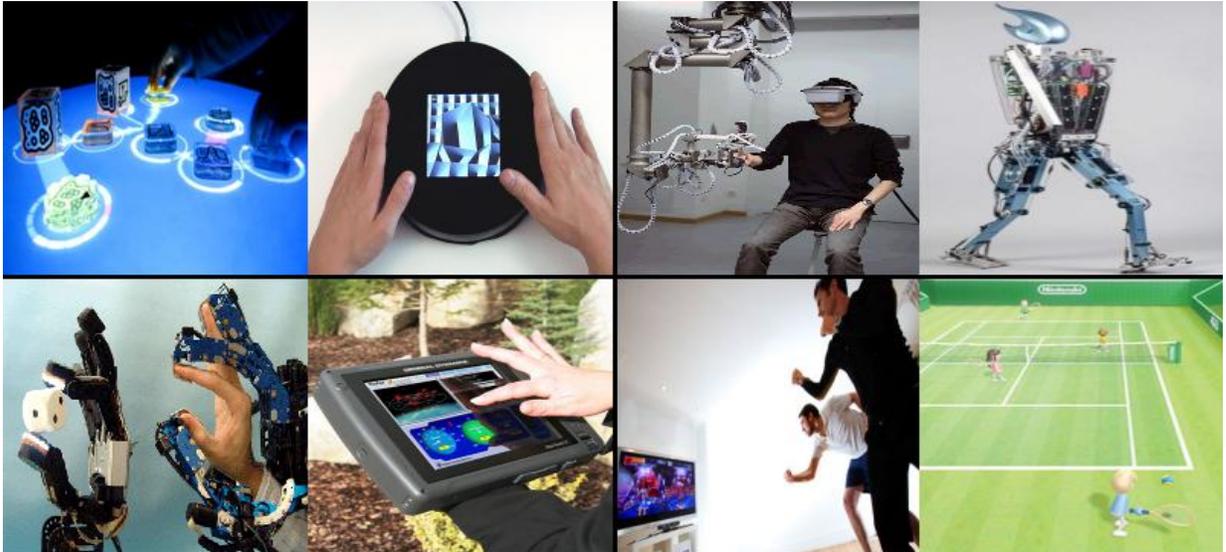

**Figure 1. A visual survey of touch-based devices that may interact with virtual worlds. Upper left frame: tangible user interfaces (TUIs). Upper right frame: biomimetic robotic devices with "lifelike" touch and motion, Lower left frame: haptic interfaces, Lower right frame: Kinect and Wii video game consoles.**

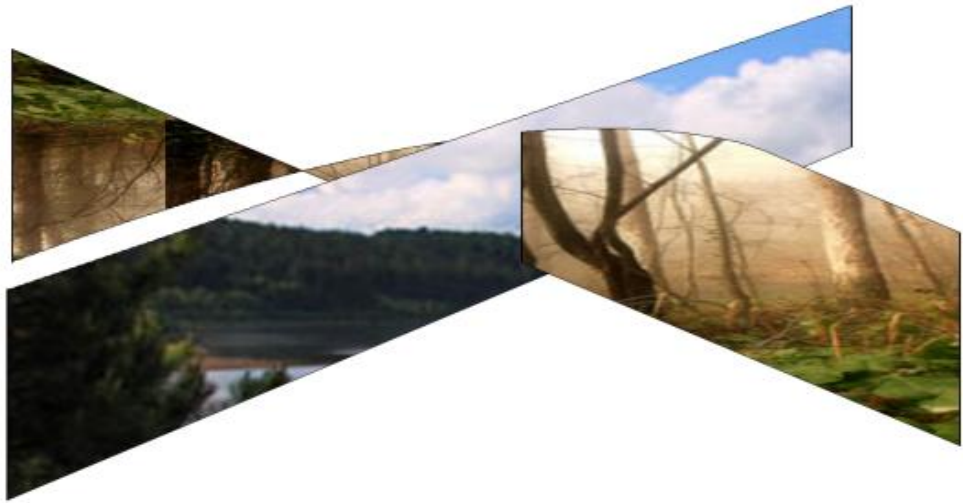

**Figure 2. A cartoon example of two orthogonal, warped dimensions, each representing virtual information. The common currency of these dimensions is a light constant, which serves to unify forces, gravity, and material properties.**

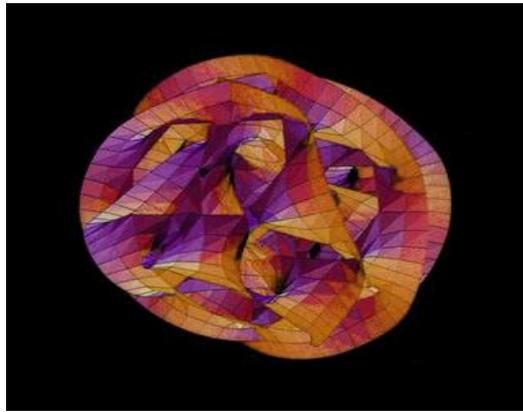

**Figure 3. An example of the Calabi-Yau manifold, an 11-dimensional topology.**

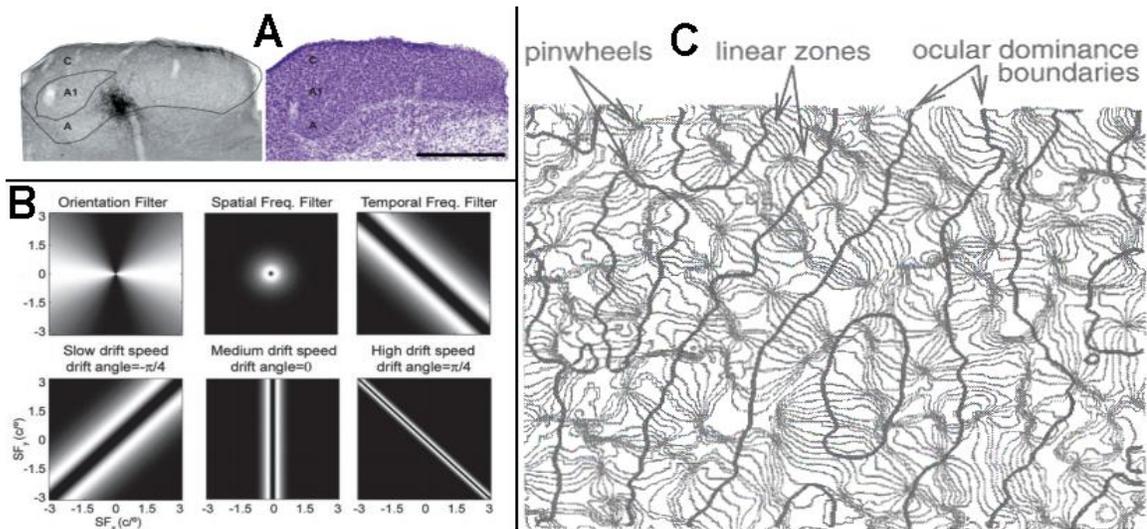

**Figure 4. Examples of neural representations, which transform sensory information into higher dimensions and representational subspaces. A) ocular dominance in the thalamus [34], separable pattern filtering response in the visual cortex [35], and color vision processing in the visual cortex [36].**

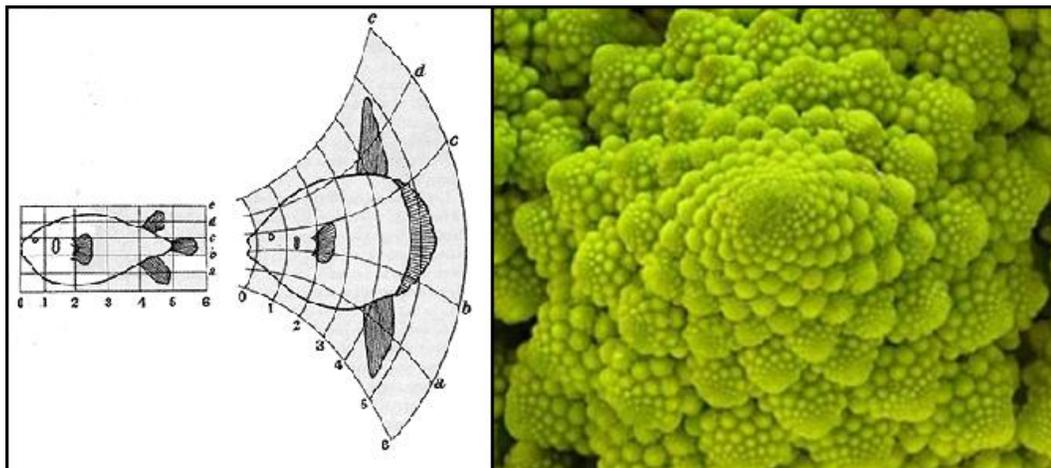

**Figure 5. Examples of allometry [42] and isometry [43]. Left: Mola fish body shape scales allometrically as it grows in size. Right: Romanesco broccoli leaves scale isometrically as it grows in size.**

# References

[1] Alcaniz, M., Lozano, J.A., and Rey, B. (2006). Technological background about virtual reality. In Cybertherapy, Chapter 10. G. Riva, C. Botella, and P. Legeron, eds. IOS Press, Amsterdam.

[2] Sherman, W.R. and Craig, A.B. (2003). Understanding Virtual Reality: interface, application, and design. Morgan-Kauffman, San Francisco.

[3] Heim, M. (1993). The Metaphysics of Virtual Reality. Oxford Press, Oxford, UK.

[4] Murray, D.J. et.al (1999). On the size-weight illusion. Perception and Psychophysics, 61, 1681-1685.

[5] Jones, L.A. and Lederman, S.J. (2006). Human Hand Function. Oxford University Press, Oxford, UK.

[6] Chemero, A. (2009). Radical embodied cognitive science. MIT Press, Cambridge, MA.

[7] Clark, A. (1997). Being there. Oxford Press, Oxford, UK.

[8] Rees, G., Kreiman, G., and Koch, C. (2002). Neural correlates of consciousness in humans. Nature Reviews Neuronscience, 3, 261-270.

[9] Wickens, C.D. and Baker, P. (1995). Cognitive Issues in Virtual Reality. In "Virtual Environments and Advanced Interface Design", W. Barfield and T.A. Furness, eds. pgs. 514-541. Oxford Press, Oxford, UK.

[10] Bronstein, A.M., Bronstein, M.M., and Kimmel, R. (2008). Numerical Geometry of Non-Rigid Shapes. Springer, Berlin.

[11] Norman, D.A. (2002). The design of everyday things. Basic Books, New York.

[12] Gabor, H.T. (1998). Geometry of Digital Spaces. Birkhauser, Boston.

[13] Zhao, Q. Ten Scientific Problems in Virtual Reality. Communications of the ACM, 54(2), 116-118.

[14] Otaduy, M.A. and Lin, M.C. (2006). High-fidelity Haptic Rendering. Morgan and Claypool, San Rafael, CA.

[15] McCarthy, W. (2003). Hacking Matter: levitating chairs, quantum mirages, and the infinite weirdness of programmable atoms. Basic Books.

[16] Avila, R., Basogan, C., Massie, T., Staples, D., Ruspini, D., Salisbury, K., and Taylor, R. (1999). Haptics: from basic principles to advanced applications. SIGGRAPH Lecture Notes, No. 38.

[17] The Computational Neurobiology of Reaching and Pointing. MIT Press, Cambridge, MA.

[18] MacKensie, I.S. and Buxton, W. (1992). Extending Fitts' Law to two-dimensional tasks. Proceedings of SIGCHI, ACM Press, 219-226.

[19] Accot, J. and Zhai, S. (1997). Beyond Fitts' Law: Models for Trajectory-Based HCI Tasks. Proceedings of SIGCHI, ACM Press, 295-302.

[20] McLaughlin, M.L., Hespanha, J.P., Sukhatme, G.S. (2002). Touch in Virtual Environments: haptics and the design of interactive systems. Prentice-Hall, Upper Saddle River, NJ.

[21] Jagacinski, R.J. and Flach, J.M. (2002). Quantitative Approaches To Modeling Performance. CRC Press, Boca Roton, FL.

[22] Mendoza, C.A., Laugier, C. (2001). Realistic haptic rendering for highly deformable virtual objects. Proceedings of the IEEE Virtual Reality, 17(17), 264-269.

[23] Dan, N. (1996). Time-dependent effects in surface forces. Current Opinion in Colloid and Interface Science, 1(1), 48-52.

[24] Baraff, D. (2001). Physically-based modeling: implicit methods for differential equations. Course Notes, SIGGRAPH, ACM Press, New York.

[25] Petkov, V. (2010). Minkowski Spacetime: A Hundred Years Later. Fundamental Theories of Physics, Volume 165. Springer, Berlin.

[26] Mignani, R. and Cardone, F. (2007). Deformed Spacetimes: geometrizing interactions in four and five dimensions. Springer, Berlin.

[27] Greene, B. (1999). The Elegant Universe: superstrings, hidden dimensions, and the quest for the ultimate theory. Vintage Books.

[28] Reichenbach, H. (1958). The Philosophy of Space and Time. Dover Publications, New York.

[29] Ellis, G.F.R., Williams, R.M. (1988). Flat and Curved Space-Times. Oxford Press, Oxford, UK.

[30] Hermann, E., Raffin, B., and Faure, F. (2009). Interactive physical simulation on multicore architectures. Eurographics Workshop on Parallel Graphics and Visualization.

[31] Thomaszewski, B., Pabst, S., and Blochinger, W. (2008). Parallel techniques for physically-based simulation on multi-core processor architectures. Computers and Graphics, 32, 25-40.

[32] Stein, B.E. and Meredith, M.A. (1993). Merging of the Senses. MIT Press, Cambridge, MA.

[33] Anastasio, T.J. (2010). Tutorial on Neural Systems Modeling. Sinauer, Sunderland, MA

[34] Baker, T.I. and Issa, N.P. (2005). Cortical maps of seperable tuning properties predict population responses responses to complex visual stimuli. Journal of Neurophysiology, 94, 775-787.

[35] Crowley, J.C. and Katz, L.C. (2000). Early Development of Ocular Dominance Columns. Science, 290, 1321-1324.

[36] Dayan, P. and Abbott, L.F. (2001). Theoretical Neuroscience: computational and mathematical modeling of neural systems. MIT Press, Cambridge, MA.

[37] Biewener, A.A. (1983). Allometry of quadrupedal locomotion: The scaling of duty factor, bone curvature and limb orientation to body size. Journal of Experimental Biology, 105, 147–171.

[38] Bar-Cohen, Y. (2005). Biomimetics: biologically-inspired technologies. Taylor and Francis, New York.

[39] Sanchez-Vives, M.V. and Slater, M. (2005). From Presence to Consciousness through Virtual Reality. Nature Reviews Neuroscience, 6, 332-339.

[40] Holmes, N.P. and Spence, C. (2004). The body schema and multisensory representations of peripersonal space. Cognitive Processes, 5(2), 94-105.

[41] Spence, C., Pavani, F., Maravita, A., and Holmes, N. (2004). Multisensory contributions to the 3-D representation of visuotactile peripersonal space in
humans: evidence from cross-modal congruency task. Journal of Physiology: Paris, 98(1-3), 171-189.

[42] Thompson, D. (1917). On growth and form. Cambridge University Press, Cambridge, UK.

[43] Flake, G.W. (2000). Computational beauty of nature: computer explorations of fractals, chaos, complex systems, and adaptation. MIT Press, Cambridge, MA.

[44] Binder, K. (1987). Theory of first-order phase transitions. Reports on Progress in Physics, 50(7), 783.

[45] Winfree, A. (1980). Geometry of Biological Time. Springer, Berlin.

[46] Hord, R.M. (1999). Understanding Parallel Computing. IEEE Press, Piscataway, NJ.

[47] Scott, L.R., Clark, T., and Bagheri, B. (2005). Scientific Parallel Computing. Princeton University Press, Princeton, NJ.